\documentclass[superscriptaddress,twocolumn,prl]{revtex4-2}
\usepackage{graphicx}
\usepackage{dcolumn}
\usepackage{bm}
\usepackage{color}
\usepackage{ulem}
\usepackage{amsmath}
\usepackage{amsfonts}
\usepackage{amssymb}
\usepackage{appendix}
\usepackage{array}
\usepackage{tikz}
\usepackage[compat=1.1.0]{tikz-feynman}
\usepackage{changes}
\usepackage{comment}
\usepackage{multirow}
\usepackage[colorlinks=True,linkcolor=blue,anchorcolor=red,citecolor=blue,CJKbookmarks=true]{hyperref}

\begin{document}

\title{$\frac{5}{2}$ fractional quantum Hall state in GaAs with Landau level mixing}
\author{Wenchen Luo}\email{luo.wenchen@csu.edu.cn}
\affiliation{School of Physics, Central South University, Changsha, China
410083}
\author{Muaath Abdulwahab}
\affiliation{School of Physics, Central South University, Changsha, China
410083}
\author{Xiang Liu}
\affiliation{School of Physics, Central South University, Changsha, China
410083}
\author{Hao Wang}
\email{wangh@sustech.edu.cn}
\affiliation{Shenzhen Institute for Quantum Science and Engineering,
Southern University of Science and Technology, Shenzhen, China 518055}
\affiliation{International Quantum Academy, Shenzhen, China 518048}
\affiliation{Guangdong Provincial Key Laboratory of Quantum Science and
Engineering, Southern University of Science and Technology, Shenzhen, China
518055}
\date{\today }

\begin{abstract}

The Landau level mixing is the key in understanding the mysterious $5/2$ fractional quantum Hall effect in GaAs quantum well. Theoretical calculations with and without Landau level mixing show striking differences. However, the way to deal with the considerable strong Landau level mixing in GaAs is still unsatisfactory. We develop a method combining the screening and the perturbation theories to study the nature of the $5/2$ fractional quantum Hall effect in GaAs efficiently. The screening which has been succeed in explaining ZnO systems integrates out the low-energy Landau levels close to the related Landau level, while the other high-energy Landau levels are integrated out by the perturbation theory. We find that the ground states still hold the quasi-triplet degeneracy which implies the Pfaffian nature of the system. Furthermore, the particle-hole symmetry is only weakly violated since the particle-hole parity is close to unity. We propose that the ground state in the finite-size calculations can be approximated as a variational superposition of the Pfaffian and anit-Pfaffian states. In the experimental environment the symmetrized Pfaffian component is dominant, corresponding a thermal conductance around $2.5$ quanta can be understood consequently.

\end{abstract}

\maketitle

Since the discovery of the even-denominator fractional quantum Hall effect (FQHE) in 1989, many efforts have been made to understand this extraordinary phenomena \cite{odfqhe,chapter}. This topic is constantly attractive, not only its non-Abelian excitation is supposed to be useful in topological quantum computation \cite{nonab}, but also it enriches the knowledge of topological order of strong-correlated state \cite{review}. However, thus far, the nature of the $5/2$ FQHE is still in puzzle. To explore this phenomenon, people have proposed several trial wave functions \cite{trialwf,mr,son,apf}, but none of them could well explain either the related experiments or the numerical results. Among these trial wave functions, Moore-Read Pfaffian and its particle-hole (PH) conjugate, anti-Pfaffian, states are the most likely candidates capturing the nature of the $5/2$ FQHE \cite{mr,apf}. The two states are topologically different and can be distinguished by some topological quantities, such as the spherical shift, or edge currents. In numerical studies without Landau level mixing (LLM), the ground state has been found to be PH symmetric and have the same overlap with the Pfaffian or with the anti-Pfaffian state. However, once the LLM is taken into consideration, the PH symmetry of the system may be broken due to the emerging many-body interactions. It seems that the understanding of the $5/2$ FQHE must rely on properly addressing LLM.

In order to calibrate the strength of LLM, ones define a parameter $\kappa$ by the ratio of Coulomb energy $e^2/\epsilon \ell$ to the Landau level (LL) gap $\hbar \omega_c$ with the dielectric constant $\epsilon$ and the magnetic length $\ell=\sqrt{\hbar/e B}$ for the magnetic field $B$.  In semiconductor quantum wells the cyclotron frequency is $\omega_c = eB/m^*$ with the effective mass $m^*$, thus $\kappa \propto m^*/\epsilon \sqrt{B}$. For the $5/2$ FQHE with weak LLM as $\kappa \ll 1$, numerical studies with the perturbation theory adopted three-body corrections up to the first order of $\kappa$, results of which showed the ground state to be PH non-symmetric. When the LLM is strong enough, such as $\kappa > 5$ in ZnO, the PH symmetry could be recovered by the renormalized Coulomb interaction due to screening in the linear response theory \cite{linearresponse}. The ground state there has been found to be the symmetrized Pfaffian state \cite{spf} with PH symmetry, which is the superposition of the Pfaffian and the anti-Pfaffian states with equal weights \cite{falson1,falson2,joe_more,luo1,luo2021,luo2022}.

However, in the most common GaAs quantum well with the moderate LLM of $\kappa \gtrsim 1$, the ground state of $5/2$ FQHE still causes much controversy \cite{argue,chapter}. One practical way of recognizing the nature of this incompressible state and distinguishing different candidate states is to measure the thermal conductance of the system. The existence of the half thermal conductance quanta suggests the Majorana edge mode, which rules out those trial wave functions with Abelian excitations. The remaining non-Abelian candidates can be judged by different thermal channels due to their topological difference. Recent experiments seem to rule out either of the two most popular candidates, Pfaffian and anti-Pfaffian states \cite{thermal1,thermal2}. Another candidate, the PH-Pfaffian state \cite{son}, which is constructed by the $s$-wave pairing of the composite fermion with PH symmetry is in doubt, due to the energy unfavorable in numerical calculations \cite{pfenergy}. Although many proposals are discussed to explain the experiments \cite{thermaldiscuss}, it seems far from reaching a conclusion.

The $5/2$ FQHE in GaAs is difficult to precisely calculate with its moderate LLM. On the one hand, the perturbation theory \cite{perturbation,perturbation2} may not be accurate while different truncations on the higher orders lead different results \cite{loic}. On the other hand, the screening approach suitable for larger $\kappa$ cases omits PH symmetry breaking mechanism, which could be important in the moderate $\kappa$ case. We now adapt a strategy of embedding the screening theory into the perturbation theory, which integrates the advantages of the two theories, and circumvents the difficulty of moderate $\kappa$.

We consider a two-dimensional electron gas in GaAs trapped in an infinite square well \cite{peterson} in the $z$ direction. A LL can then be labeled by $(m, n, \sigma)$ where $m=1,2\ldots$ is the band index, $n=0,1\ldots$ is the LL index and $\sigma=\pm$ is the spin index. For 5/2 FQHE, the studied LL is labeled by $(1,1,+)$. We adopt experimental parameters from the conventional GaAs quantum well with the effective mass of electron $m^*=0.067m_e$, the dielectric constant $\epsilon=12.9$, the quantum well width $W=30 \sim 40$ nm and the magnetic field $B=3 \sim 6$ T, which leads to $\kappa \sim 1$. We note some lowest LLs in the second band can be very close to our studied LL, e.g., LLs $(2,0,\pm)$ gapped from $(1,1,+)$ by only $0.194 e^2/\epsilon \ell$ with $B=5\,$T and $W=40$ nm. Thus, we should include these levels in the screening process.

In the following, we describe our algorithm with three steps in details. Firstly, we classify all the LLs into three sets: the set $\mathcal{A}=\{(1,1,+)\}$ only has the studied LL; the set $\mathcal{B}$ contains all the adjacent LLs of the studied LL with the energy gap less than $e^2/\epsilon \ell$ from the studied LL. In our calculations, we have $\mathcal{B}=\{(1,0,\pm),(1,2,\pm), (1,1,-)$, $(2,0,\pm) \}$; and the set $\mathcal{C}$ includes all other LLs. In practice, we truncate LLs with energy more than $(10+E_{1,1,+})e^2/\epsilon \ell$. The truncation with a larger energy threshold does not affect the numerical results. Secondly, we integrate out the LLs in set $\mathcal{B}$ by using the screening theory. Note that the virtual processes in the screening between the studied LL $(1,1,+)$ and the LLs in set $\mathcal{C}$ should be excluded since the related LLM effects will be considered in the perturbation theory. The Coulomb interaction outside of the screening set is then  renormalized by a screened dielectric function. The LLs in set $\mathcal{C}$ are at least $e^2/\epsilon \ell$ away from the studied LL, which guarantees that the effective LLM parameter is reduced to $\kappa_{\text{eff}} < 1$. Thirdly, by using the perturbation theory, all other LLs except the studied LL are integrated out so that a one-LL effective Hamiltonian is obtained. Our approach combines the screening and perturbation theories. Particularly, it keeps the three-body interaction that may break the PH symmetry in case where $\kappa$ is not large enough to restore this symmetry completely. Thus, this method allows one to effectively deal with arbitrary LLM systems, especially the GaAs quantum well system.

The screened Coulomb interaction is renormalized by the dielectric function \cite{luo2021,luo2022} in form of
\begin{equation}
\epsilon_s\left(\mathbf{q,}q^{}_z\right) =1+\frac{4\pi e^2}{\left(
q^2+q_z^2\right)\epsilon}\chi^0\left(\mathbf{q,} q_z\right)
\end{equation}
with the in-plane momentum $\mathbf{q}$ and out-of-plane momentum $q_z$. The static retarded density-density response function is calculated by the random phase approximation (RPA) as
\begin{equation}
\lim_{\omega \rightarrow 0} \chi^{0}\left(\mathbf{q} ,q_{z}, \omega \right)    =
\overline{\sum_{\mathbf{1,2},\sigma } } \frac{\left\vert
G_{\mathbf{1,2}}\left( \mathbf{q},q_{z}\right)
\right\vert ^{2}} {2\pi \ell ^{2}W}
\frac{\nu _{\mathbf{1} ,\sigma }-\nu _{\mathbf{2},\sigma }}{E_{ \mathbf{2}
,\sigma}
-E_{ \mathbf{1} ,\sigma }},
\label{chi}
\end{equation}
where the bold numbers $\mathbf{i=1,2}$ are the short notations of the LLs with $(m_{\mathbf{i}},n_{\mathbf{i}})$. $E_{\mathbf{i},\sigma } $ and $\nu_{\mathbf{i},\sigma }$ are the kinetic energy and the noninteracting filling factor of the LL $(m_{\mathbf{i}}, n_{\mathbf{i}}, \sigma)$, respectively. The form factor $G$ has its detailed expression in the supplementary material (SM) \cite{sm}. The bar over the sum means that only the virtual precesses containing at least one LL in set B are $included$ while all other processes will be handling in the following perturbation treatment.

\begin{figure}[htbp]
\centering
\includegraphics[width=8.5cm]{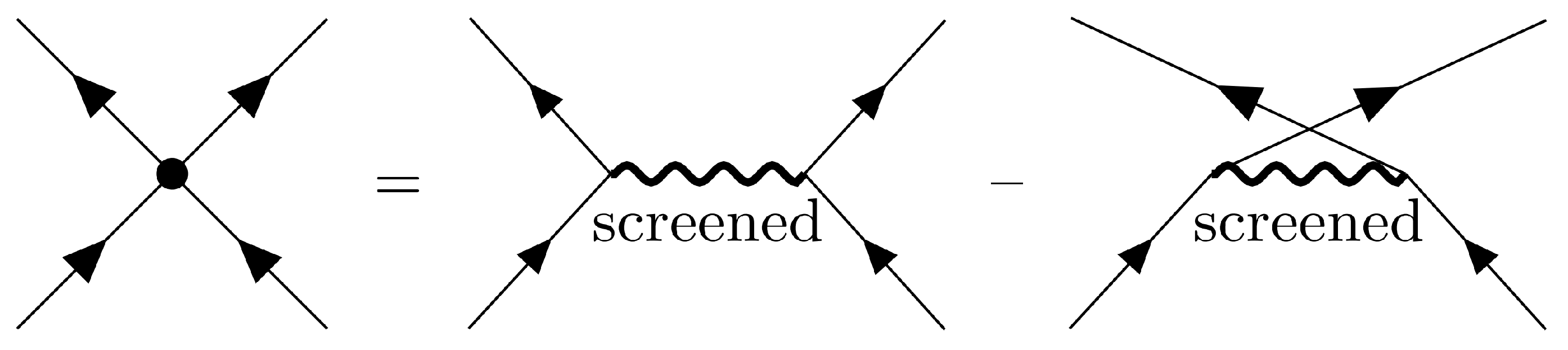}
\caption{The Feynman diagram of the Coulomb interaction vertex. The bare Coulomb potential is replaced by the screened Coulomb potential calculated in the RPA, as the dressed wave lines shown in the figure. }
\label{feynmandiagram}
\end{figure}

In the Feynman diagrams \cite{perturbation2} of the perturbation theory, all the bare vertices in the two-body corrections (Bardeen–Cooper–Schrieffer ($BCS$), zero sound ($ZS$) and $ZS'$ diagrams) and the three-body corrections need to be replaced by the screened ones, as shown in Fig. \ref{feynmandiagram}.
These screened Coulomb vertices have the form of
$V_{\mathbf{12;34}}^{\alpha' \beta';\beta \alpha}=\delta_{\alpha',\alpha} \delta_{\beta', \beta}V^{s,\mathbf{1,2;3,4}}_{i_1,i_2;i_3,i_4}-\delta_{\alpha',\beta}\delta_{\beta',\alpha}V^{s,\mathbf{2,1;3,4}}_{i_2,i_1;i_3,i_4}$ , where $\alpha,\beta$ are spin indices and $i_{j}$ are orbit indices for the $j-$th LL which are the guiding center indices in the Landau gauge. The detailed formulas for these terms are expressed in the SM \cite{sm}. With these vertices we can write down the effective Hamiltonian \cite{perturbation2} in form of $H_\text{eff}= H_0 + BCS +ZS +ZS' + H_\text{3b}$, where all correction terms are smaller quantities comparing with the LL gap. In our case, the $ZS$ and $ZS'$ diagrams would be vanished if all the LLs below the Fermi level have been included in the screening set $\mathcal{B}$. The detailed expressions for the effective Hamiltonian are provided in the SM \cite{sm}.

\begin{figure}[htbp]
\centering
\includegraphics[width=9.1cm]{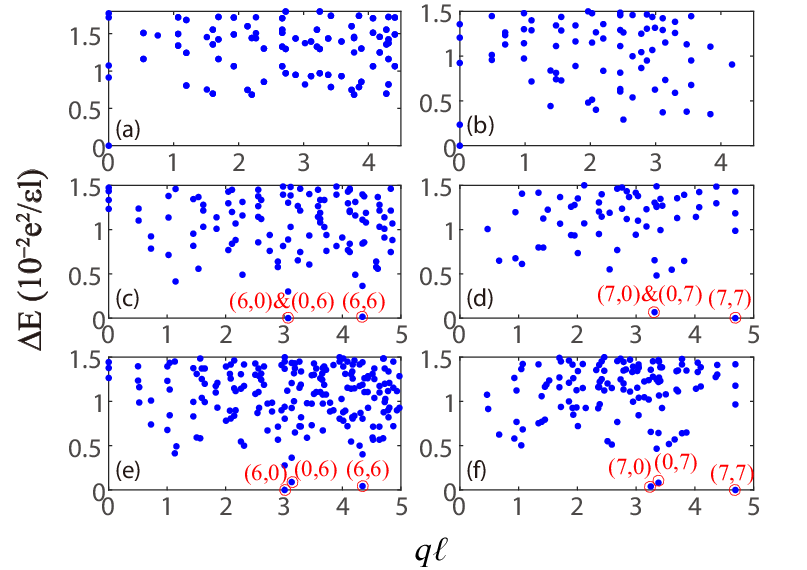}
\caption{(Color online) The low-energy excitation spectra of systems ($B=5$ T and $W=40$ nm) with different electrons and different unit cell geometries: (a) $N_e=11$, $\tau=i$; (b) $N_e=13$, $\tau=i$; (c) $N_e=12$, $\tau=i$; (d) $N_e=14$, $\tau=i$; (e) $N_e=12$, $\tau=0.96i$; (f) $N_e=14$, $\tau=0.96i$.
The ground states for odd-electron systems are all located at $q=0$ while the ground states of even-electron systems are quasi-triple-degenerate at the three characteristic pseudomomentum sectors marked by circles in (c) to (f).
}
\label{fig2}
\end{figure}

We carry out numerical studies with $H_\text{eff}$ on torus \cite{haldane} by exact diagonalization for systems with the electron number up to $N_e=14$. The cell geometry of the torus is parameterized by $\tau=|\tau| e^{i \theta}$ where $|\tau|$ is the aspect ratio and $\theta$ is the aspect angle. We do not use the spherical geometry since the spherical shift cannot be determined in advance. The low-lying energy spectra for several example systems are shown in Fig. \ref{fig2}. For odd-electron systems, the spectra demonstrate a gapped ground state locating at $\mathbf{q}=0$ as in Figs. \ref{fig2}(a) and (b), indicating stable incompressibility of the $5/2$ FQHE \cite{luo1,luo2021,luo2022}.

For even-electron systems, there exist quasi-triplet ground states locating at the pseudomomentum sectors $(N_e/2,0)$, $(0,N_e/2)$ and $(N_e/2, N_e/2)$ as shown in Figs. \ref{fig2}(c)-(f), which are exact locations of Pfaffian (or anti-Pfaffian) states on torus. We have checked the systems with different numbers of electrons and with different cell geometries around $|\tau| \sim 1$, where the triplet ground state feature is general. When the cell aspect ratio is extremely large or small, the quasi-triplet degeneracy is lift due to the anisotropic response on the screened dielectric function. In the following discussion, we will mainly show our numerical results with a square cell ($\tau = i$), where the two states on $(N_e/2,0)$ and $(0,N_e/2)$ sectors are exact degenerate due to the cell symmetry. The experimental parameters for these demonstrating systems are $B=5\,$T and $W=40$ nm.

The spectral feature of triplet ground states at characteristic momenta implies that these ground states might topologically relate to the Pfaffian-like state. To further explore the topological nature of the ground states, we calculate their Hall viscosity in the geometry parameter space \cite{vis,luo2022}, which is directly related to the topological shift in the spherical geometry \cite{wen,vis}. However, we find the calculated values cannot converge to a fixed number. Instead, they fluctuate randomly around different parameter $\tau$ corners and vary with different electron numbers. The lack of a meaningful viscosity value on torus suggests that the ground state of the system could not be described solely by the Pfaffian or anit-Pfaffian state.

Since the three-body perturbation in our effective Hamiltonian has strictly broken PH symmetry, we tend to exam the PH symmetry of the calculated ground states using PH parity which is defined by the absolute value of the overlap between the state and its PH conjugation. Numerical results for example systems with different even-numbers of electrons are collected In Table \ref{table1}. It is clear that the PH parity of the ground state is less than but very close to unity for all the cases. The separated calculations with different cell geometries provide with the similar values. This general result indicates that the PH symmetry of the system is only \textit{weakly} broken by the LLM. The same conclusion has also been drawn in the work using the fixed phase diffusion Monte Carlo method \cite{fpDMC}. It has been argued that the PH symmetry of the system can be restored gradually from the less $\kappa$ case when $\kappa > 0.6$ \cite{loic}. When $\kappa \gg 1$, the LLM is strong enough so that the PH symmetry breaker, 3-body interaction, is quite small as the multiple of two screened Coulomb vertices, thus negligible. The effective interaction then can be approximately described by the screened Coulomb interaction only. Its ground state is PH symmetric and can be described by a symmetrized Pfaffian state\cite{falson2,luo2022}. However, in our case with the conventional GaAs quantum well, a moderate $\kappa \sim 1$ appears not large enough to fully restore the PH symmetry.

\begin{table*}
\centering
\renewcommand\arraystretch{1.21}
\begin{tabular}{c|cccccc}
\hline
 \multicolumn{2}{c}{} & $N_e=6$ & $N_e=8$ & $N_e=10$ & $N_e=12$ & $N_e=14$ \\
\hline
 \multirow{4}*{\rotatebox{90}{\scriptsize{$(0,N_e/2 )$}}} & PH parity  & 0.994 & 0.998 & 0.986 & 0.840 & 0.958 \\
&$(r,\theta) \rightarrow \langle GS | trial \rangle$
& $(0.7,0) \rightarrow 0.939$  & $(0.88,\frac{3\pi}{2}) \rightarrow 0.893$
& $(0.87,0) \rightarrow 0.854$ & $(0.44,\pi) \rightarrow 0.927$
& $(0.7,\frac{\pi}{2}) \rightarrow 0.900$ \\
 &  $\langle GS | Pf \rangle$ & 0.876 & 0.762 & 0.421 & 0.545 & 0.595 \\
&$\langle GS | APf \rangle$ & 0.908 & 0.793 & 0.555 & 0.866 & 0.764 \\
&$\langle GS | SPf \rangle$ & 0.938 & 0.892 & 0.850 & 0.889 & 0.890 \\
\hline
\multirow{4}*{\rotatebox{90}{\scriptsize{$(N_e/2,N_e/2 )$}}} & PH parity  & 0.982 & 0.999 & 0.849 & 0.897 & 0.988 \\
& $(r,\theta) \rightarrow \langle GS | trial \rangle$ \,
& $(0.76,0) \rightarrow 0.999$  & $(0.92,\frac{3\pi}{2}) \rightarrow 0.933$  \,
& $(0.56,\pi) \rightarrow 0.909$ & $(0.48,\pi) \rightarrow 0.899$ \,
& $(0.76,\frac{\pi}{2}) \rightarrow 0.878$ \, \\
& $\langle GS | Pf \rangle$ & 0.768 & 0.860 & 0.386 & 0.590 &
0.714 \\
& $\langle GS | APf \rangle$ & 0.874 & 0.873 & 0.785 & 0.838 &
0.787 \\
&$\langle GS | SPf \rangle$ & 0.995 & 0.933 & 0.868 & 0.873 & 0.876 \\
\hline
\end{tabular}
\caption{The PH parity and wave function overlaps of the ground states with different trial wave functions at different pseudomonenta and in different even-number electron systems.}\label{table1}
\end{table*}

Based on the Pfaffian-like spectral feature, as well as the broken PH symmetry and uncertain Hall viscosity of the ground states, we propose a trial wave function composed by a superposition of the Pfaffian and the anti-Pfaffian states but allowing unequal weights,
\begin{equation} \label{trial}
| trial \rangle = C \left(r |Pf \rangle + e^{i \theta} |APf \rangle \right),
\end{equation}
where $C$ is the normalization factor and $|Pf \rangle$ ($|APf \rangle$) represents the Pfaffian (anti-Pfaffian) state. The variational parameters $r$ and $\theta$ are real numbers, standing for the relative ratio and phase between the two components, respectively. These parameters can be numerically determined by maximizing the wave function overlap with the calculated ground state.

We then show the maximum overlap between the optimized trial wave function $| trial \rangle$ and the ground state $|GS \rangle$ for several example systems with different numbers of electrons in Table \ref{table1}. They all present a large value ($ \ge 0.854$) close to unity. In comparison, we also provide the wave function overlap between the ground state and a single Pfaffian state $|Pf \rangle$ or anti-Pfaffian state $|APf \rangle$. With a general larger overlap, our trial wave function suggests a better description to the ground state than these two candidates. This can be further confirmed by comparing the entanglement spectra (ES) \cite{li} of these states on torus \cite{tower} as shown in the SM \cite{sm}. Meanwhile, we note the overlap with the anti-Pfaffian is always larger than the Pfaffian, indicating the anti-Pfaffian component is more in favor in the ground state. This is also evidenced by the ratios $r$ in optimized wave functions all less than unity. Calculations with a wider range of parameters $B\in [3,6]$ T and $W\in[30,40]$ nm show similar results \cite{sm}. In our numerical studies, the optimal value of the ratio $r$ varies with the number of electrons, unit cell geometries, and with experimental parameters. Therefore, the form of the wave function describing the ground state of the system is variational instead a fixed one. This partially explains why we are lack of a certain viscosity value in previous calculations. 

Considering the nearly PH symmetric property, we exam the ground state by calculating its wave function overlap with the symmetrized Pfaffian state $|SPf \rangle = |Pf \rangle + |APf \rangle$. As shown in Table \ref{table1}, this overlap remains high and is very close to that of the optimized trial wave function. The universal agreement suggests that the dominant component of the ground state can be uniformly described by this PH symmetrized state even though the whole ground state is described as a variational wave function.

It is noteworthy that the spherical ES of the PH symmetrized state \cite{luo2022} shares the similar low-level structure as that of the PH-Pfaffian state \cite{PHPfaffian_ES}: a two-side structure with the Pfaffian-featured level-counting patterns of $1,1,3,5 \ldots$ and $1,2,4,7\ldots$ from either side. These peculiar sequences are related to certain edge excitation modes of the confirm field theory, corresponding the PH-Pfaffian state a thermal conductance $2.5\kappa_0$ with $\kappa_0$ being the conductance quanta. The topological commonality on the ES suggests the symmetrized Pfaffian state could hold the same thermal conductance. Therefore, the $5/2$ FQHE in ZnO where the ground state is described as the symmetrized Pfaffian due to the strong LLM is predicted to carry the thermal conductance $2.5\kappa_0$, and for the $5/2$ FQHE in current GaAs case where the macroscopic property should be determined by the absolutely dominant symmetrized Pfaffian component, we would expect the thermal conductance around $2.5 \kappa_0$.

In principle, our method can be extended to study more general cases with a full range of $\kappa$ if we adjust the sets $\mathcal{B}$ and $\mathcal{C}$. Besides, we have studied several scenarios where the sets $\mathcal{B}$ and $\mathcal{C}$ are changed artificially. It shows that the variational trial wave function still suitable in all these cases, presenting higher overlaps with their ground states, though the PH symmetry and the favorite on either Pfaffian or anti-Pfaffian component of the ground states could vary in different scenarios. The details of these additional studies are shown in the SM \cite{sm}.

To summarize, we have developed a renormalization method to take into account the LLM effect for the $5/2$ FQHE, especially working in the region with $\kappa \gtrsim 1$.  This algorithm integrates out the LLs close to the studied LL by screening method in RPA and deals with other LLs by the perturbation theory. This allows it keep the information of breaking PH symmetry by the 3-body interactions and evades the difficulty of large $\kappa$. The according numerical studies show the ground state of the system can be reliably described by a variational trial wave function expressed in Eq. (\ref{trial}) in different experimental environments. In GaAs with $\kappa \gtrsim 1$, the ground state is almost PH symmetric, leaving its predominant part to be represented as a symmetric superposition of the Pfaffian and anti-Pfaffian components. The measured thermal conductance around $2.5\kappa_0$ in experiments \cite{thermal1,thermal2} is associated to this dominant symmetrized state. Accordingly, the ground state of the $5/2$ FQHE in ZnO with extremely strong LLM is expected to hold $2.5\kappa_0$ thermal conductance. We believe that this ansatz reveals the mysterious 5/2 FQHE to some extent, and can be used to explain the value of the thermal conductance in experiments.

W.L. is supported by the NSF-China under Grant No. 11804396. H.W. is supported by the Guangdong Major Project of Basic and Applied Basic Research (Grant No. 2023B0303000011). We are grateful to S. Simon for helpful discussions and to the High Performance Computing Center of Central South University for partial support of this work.

\onecolumngrid

\renewcommand\thefigure{S\arabic{figure}}
\renewcommand\thetable{S\arabic{table}}
\renewcommand{\theequation}{S\arabic{equation}}

\section{Formulas of the model}

\begin{figure}[htbp]
\centering
\includegraphics[width=14cm]{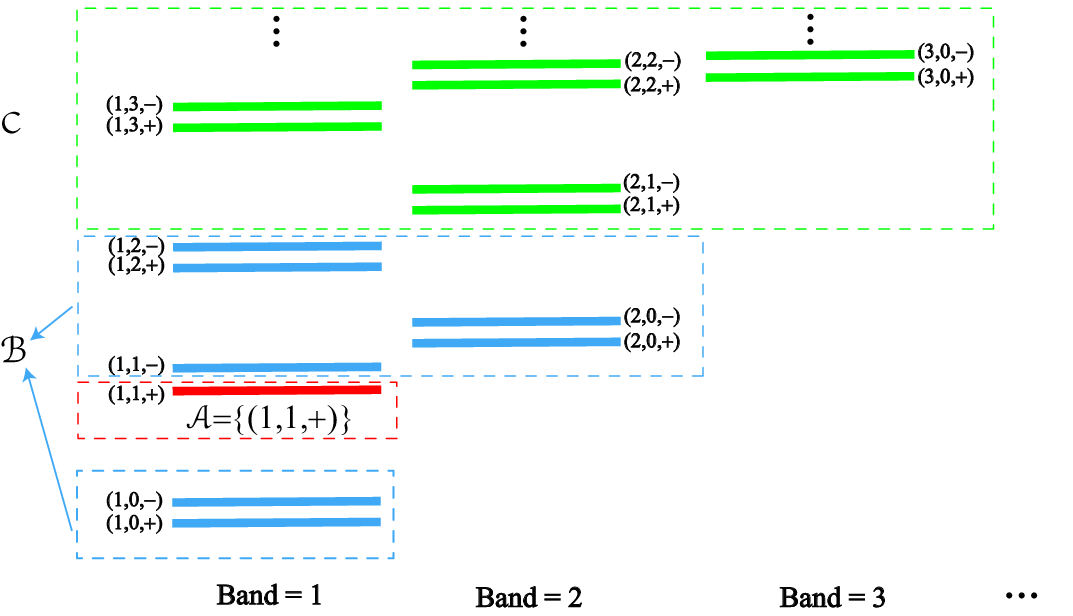}
\caption{(Color online) The classification of the LLs. The red LL is the set
$\mathcal{A}=\{(1,1,+)\}$. Blue boxes include all LLs in set $\mathcal{B}$
which are integrated out by screening. The green box includes all LLs
in set $\mathcal{C}$ which will be integrated out by the perturbation theory.
}\label{figS1}
\end{figure}

As shown in the main text, the dielectric function is given by
\begin{equation}
\epsilon^{}_s\left(\mathbf{q,}q^{}_z\right) =1+\frac{4\pi e^2}{\left(
q^2+q_z^2\right)\epsilon}\chi^0\left(\mathbf{q,} q^{}_z\right),
\end{equation}
where the density-density response function is defined by
\begin{equation}
\chi(\mathbf{q},q_z, \iota)=-\frac{1}{\hbar S W}\langle T_\iota
\delta n(\mathbf{q},q_z,\iota) \delta n(\mathbf{-q},-q_z,0) \rangle
\end{equation}
with the time ordering operator $T_\iota$, system area $S$, the width of the
well $W$, and the density operator $ n(\mathbf{q},q_z)$. The noninteracting bubble is calculated by the static retarded density-density
response function in the RPA,
\begin{equation}
\lim_{\omega \rightarrow 0} \chi^{0}\left(\mathbf{q} ,q_{z}, \omega \right)    =
\overline{\sum_{\mathbf{1,2},\sigma } } \frac{\left\vert
G_{\mathbf{1,2}}\left( \mathbf{q},q_{z}\right)
\right\vert ^{2}} {2\pi \ell ^{2}W}
\frac{\nu _{\mathbf{1} ,\sigma }-\nu _{\mathbf{2},\sigma }}{E_{ \mathbf{2}
,\sigma}
-E_{ \mathbf{1} ,\sigma }},
\label{chi}
\end{equation}%
which is the Eq. (2) in the main text. The detailed form factor is given by
\begin{equation}
G_{\mathbf{1,2}}\left( \mathbf{q},q_{z}\right)
= F_{n_{1},n_{2}}\left( -\mathbf{q}\right)
g_{m_{1},m_{2}}\left( -q_{z}\right)
\end{equation}
with the functions
\begin{eqnarray}
F_{n,n^{\prime }}\left( \mathbf{q}\right) &=& \frac{\sqrt{\min \left(
n,n^{\prime }\right) !}}{\sqrt{\max \left( n,n^{\prime }\right) !}}e^{-\frac{%
q^{2}\ell ^{2}}{4}}L_{\min \left( n,n^{\prime }\right) }^{\left\vert
n-n^{\prime }\right\vert }\left( \frac{q^{2}\ell ^{2}}{2}\right)
\left[ \frac{\mathtt{sgn}\left( n'-n \right) q_{y}\ell
+iq_{x}\ell }{\sqrt{2}}\right] ^{\left\vert n-n^{\prime }\right\vert }
\label{ffunction} \\
g_{m,m'}\left( q_{z}\right) &=& -iq_z W \left[ 1-e^{iq_{z}W }\cos \left(
m - m' \right) \pi \right]
\left[ \frac{1}{\left( m-m' \right) ^{2} \pi^2 -
q_{z}^{2}W^{2} }-\frac{1}{\left( m+ m' \right) ^{2}\pi^2-
q_{z}^{2}W^{2}}\right] .
\end{eqnarray}
The screening scheme
can be explained diagrammatically in Fig. \ref{figS1}. The virtual processes
inside of the blue levels (set $\mathcal{B}$), the virtual processes between
the blue levels and the red level $[ \mathcal{A}=\{(1,1,+)\} ]$, and the virtual
processes between the blue levels and the green levels (set $\mathcal{C}$)
are taken into account in the density-density response of the dielectric
function. In such way, the LLs in set $\mathcal{B}$ are integrated out.

In the screened Coulomb vertices, $V_{\mathbf{12;34}}
^{\alpha' \beta';\beta \alpha}=\delta_{\alpha',\alpha} \delta_{\beta', \beta}
V^{s,\mathbf{1,2;3,4}}_{i_1,i_2;i_3,i_4} -\delta_{\alpha',\beta} \delta_{\beta',\alpha}
V^{s,\mathbf{2,1;3,4}}_{i_2,i_1;i_3,i_4}$, the Coulomb interaction
matrix elements has the form in the toroidal geometry as,
\begin{eqnarray}
V^{s,\mathbf{1,2;3,4}}_{i_1,i_2;i_3,i_4} &=&
\frac{e^2}{\epsilon \ell} \frac{2}{\pi N^{}_s}\sum_{\mathbf{q\neq 0}}
\delta _{i_{1},i_{4}+q_{y}\ell ^{2}}^{\prime }\delta
_{i_{2},i_{3}-q_{y}\ell ^{2}}^{\prime }e^{iq_{x} \ell^2 \left(
i_{3}-i_{1}\right) }
 F_{n_1,n_4}(\mathbf{q}) F_{n_2,n_3}(\mathbf{-q})
V^{s,z}_{m_1,m_2,m_3,m_4}(\mathbf{q})
\end{eqnarray}
with $\delta^{\prime }$ resulting from the periodic boundary conditions and the
thickness correction of the quantum well is embedded in the expression
\begin{equation}
\hspace{-0.2em} V^{s,z}_{m_1,m_2,m_3,m_4}(\mathbf{q})= \int_0^{\infty }\frac{
g_{m_1,m_4}(q_z) g_{m_2,m_3}(-q_z) dq^{}_z}{ \pi^4 \epsilon^{}_s
\left( \mathbf{q} ,q^{}_z\right) \left( q^2+q_z^2\right) \ell}.
\end{equation}

In this work, we suppose that the spin is polarized since our
previous work and most experimental results indicate that the spin of
$5/2$ FQHE is polarized. Noting all the energy gaps between the red LL
and the green LLs are larger than $e^2/\epsilon \ell$, we can integrate out the LLs in
set $\mathcal{C}$ by using the perturbation theory safely. Then the
spin-polarized effective many-body Hamiltonian within the single LL
$(1,1,+)$, up to the first-order corrections of the LLM, is given by
\begin{eqnarray}
H_\text{eff}&=&H_0+ H_{BCS}+H_{ZS}+H_{ZS'} +H_{3b}\\
&=& \sum_{1,\ldots,4} V_2 ({1,2,3,4})
c^\dag_{1}c^\dag_{2} c_{3} c_{4}
+ \sum_{1,\ldots,6} V_3 (1,2,3,4,5,6)
c^\dag_{1}c^\dag_{2} c^\dag_{3} c_{4} c_5 c_6,  \label{heff}
\end{eqnarray}
where the spin index is neglected and $1,\ldots,6$ represent the different LL orbits of the LL
$(1,1,+)$ with their corresponding guiding centers $i_{1,\ldots,6}$. $H_0$ is
the Coulomb interaction within the LL $(1,1,+)$ with the screening
correction. The interaction matrix elements $V_2$, $V_3$
are explicitly given by
\begin{eqnarray}
&& V_{2}\left( 1,2,3,4\right)  =\frac{1}{2}V^{s,1,2;3,4}_{i_1,i_2;i_3,i_4}+
\frac{1}{4}\sum_{\mathbf{x},\mathbf{x}^{\prime }}^{\prime }
V_{\mathbf{x}\mathbf{x}^{\prime
};34}^{++;++}V_{12;\mathbf{x}\mathbf{x}'}^{++;++}
\frac{\theta \left( \widetilde{E}_{\mathbf{x},+}\right) -
\theta \left( -\widetilde{E}_{\mathbf{x}', +}\right) }
{-\widetilde{E}_{\mathbf{x}',+}-\widetilde{E}_{\mathbf{x},+}}
\\
&&+\sum_{\mathbf{x},\mathbf{x}'}^{\prime }\left[
V_{\mathbf{x}2;\mathbf{x}' 4}^{++;++}V_{1\mathbf{x}' ;3\mathbf{x}}^{++;++}
\frac{\theta \left( \widetilde{E}_{\mathbf{x},+}\right)
-\theta \left( \widetilde{E}_{\mathbf{x}',+}\right) }
{\widetilde{E}_{\mathbf{x}', +}-\widetilde{E}_{\mathbf{x},+}}
+V_{\mathbf{x}2;\mathbf{x}' 4}^{-+;-+}V_{1\mathbf{x}' ;3\mathbf{x}}^{+-;+-}
\frac{\theta \left( \widetilde{E}_{\mathbf{x},-}\right)
-\theta \left( \widetilde{E}_{\mathbf{x}', -}\right) }
{\widetilde{E}_{\mathbf{x}', -}-\widetilde{E}_{\mathbf{x}, -}}\right]
\notag \\
&&+\frac{2}{3!}\sum_{x_0}\sum_{\mathbf{x}}^{\prime }\left[ \frac{\theta
\left( -\widetilde{E}_{\mathbf{x}, +}\right) }{\widetilde{E}_{\mathbf{x}, +}}
\sum_{\{2,1,x_0\}_\text{c.p.}} \sum_{\{x_0,4,3\}_\text{c.p.}}
V_{\mathbf{x}2;x_04}^{++;++}V_{1x_0;3\mathbf{x}}^{++;++}
- \frac{\theta \left( \widetilde{E}_{\mathbf{x},+}\right) }
{4 \widetilde{E}_{\mathbf{x},+}} \sum_{\{x_0,1,2\}_\text{c.p.}}
\sum_{\{3,4,x_0\}_\text{c.p.}}
V_{\mathbf{x}x_0;34}^{++;++}V_{12;\mathbf{x}x_0}^{++;++} \right],
\notag \\
&&V_{3}\left( 1,2,3,4,5,6\right) =-\frac{1}{3!}\sum_{\mathbf{x}}^{\prime
}\frac{1}{\widetilde{E}_{\mathbf{x},+}}
\sum_{\{1,2,3\}_\text{c.p.}}\sum_{\{5,6,4\}_\text{c.p.}}
V_{1\mathbf{x};56}^{++;++}V_{23;\mathbf{x}4}^{++;++} ,
\end{eqnarray}
where $x_0$ labels the guiding center $i_{x_0}$ at the LL $(1,1)$ and the bold $\mathbf{x}$ for a generic LL $(m_x,n_x)$. $\theta$ is the Heaviside step function and $\widetilde{E}_{\mathbf{x},\sigma}=E_{\mathbf{x},\sigma} - E_{1,1,+}$ is the energy difference between LL $(\mathbf{x},\sigma)$ and the Fermi surface. The symbol $\{1,2,3\}_\text{c.p.}$ represents the cyclic permutation of the three indices. The prime over the sum means that the LLs in sets $\mathcal{A}$ and $\mathcal{B}$ are excluded, i.e. $\mathbf{x,x'} \in \mathcal{C}$. $V_2$ includes the normal ordering of the three-body interaction. Our results are mainly based on exactly diagonalizing $H_\text{eff}$ in the toroidal geometry. The reason that we do not use the spherical geometry is that the spherical shift can not be determined in advance given the three-body terms break the PH symmetry.

\onecolumngrid

\section{Algorithms Comparison}
\begin{table*}[htb]
\centering
\renewcommand\arraystretch{1.4}
\begin{tabular}{c|c|ccccc}
\hline
 \multicolumn{2}{c|}{} & $N_e=6$ & $N_e=8$ & $N_e=10$ & $N_e=12$ & $N_e=14$ \\
\hline
\multirow{4}*{\rotatebox{90}{\scriptsize{$(0,N_e/2 )$}}} & PH parity  & 1 & 1 & 1 & 1 & 1 \\
&$(r,\theta) \rightarrow \langle GS | trial \rangle$
& $(1,0) \rightarrow 0.926$  & $(1,\frac{3\pi}{2}) \rightarrow 0.866$
& $(1,0) \rightarrow 0.802$ & $(1,\pi) \rightarrow 0.861$
& $(1,\frac{\pi}{2}) \rightarrow 0.860$ \\
& $\langle GS | Pf \rangle$ & 0.881 & 0.755 & 0.460 & 0.683 & 0.657 \\
&$\langle GS | APf \rangle$ & 0.881 & 0.755 & 0.460 & 0.683 & 0.657 \\
\hline
\multirow{4}*{\rotatebox{90}{\scriptsize{$(N_e/2,N_e/2 )$}}} &PH parity  & 1 & 1 & 1 & 1 & 1 \\
&$(r,\theta) \rightarrow \langle GS | trial \rangle$
& $(1,0) \rightarrow 0.999$  & $(1,\frac{3\pi}{2}) \rightarrow 0.909$
& $(1,\pi) \rightarrow 0.936$ & $(1,\pi) \rightarrow 0.716$
& $(1,\frac{3\pi}{2}) \rightarrow 0.756$ \\
&$\langle GS | Pf \rangle$ & 0.824 & 0.844 & 0.564 & 0.586 &
0.648 \\
& $\langle GS | APf \rangle$ & 0.824 & 0.844 & 0.564 & 0.586 &
0.648 \\
\hline
\end{tabular}
\caption{The ground states here are calculated by using screening
theory only. The overlaps between ground states and different trial wave
functions and the PH parity of the ground states.}\label{table3}
\end{table*}

\begin{table*}[htb]
\centering
\renewcommand\arraystretch{1.4}
\begin{tabular}{c|c|ccccc}
\hline
  \multicolumn{2}{c|}{} & $N_e=6$ & $N_e=8$ & $N_e=10$ & $N_e=12$ & $N_e=14$ \\
\hline
\multirow{4}*{\rotatebox{90}{\scriptsize{$(0,N_e/2 )$}}} &PH parity  & 0.924 & 0.911 & 0.226 & 0.329 & 0.439 \\
&$(r,\theta) \rightarrow \langle GS | trial \rangle$
& $(4.06,0) \rightarrow 0.989$  & $(2.35,\frac{3\pi}{2}) \rightarrow 0.947$
& $(19.9,0) \rightarrow 0.971$ & $(25.3,\pi) \rightarrow 0.968$
& $(6.94,\frac{\pi}{2}) \rightarrow 0.969$ \\
& $\langle GS | Pf \rangle$ & 0.982 & 0.907 & 0.970 & 0.966 & 0.959 \\
& $\langle GS | APf \rangle$ & 0.866 & 0.702 & 0.287 & 0.286 & 0.291 \\
\hline
\multirow{4}*{\rotatebox{90}{\scriptsize{$(N_e/2,N_e/2 )$}}} & PH parity & 0.796 & 0.958 & 0.242 & 0.313 & 0.745 \\
& $(r,\theta) \rightarrow \langle GS | trial \rangle$
& $(2.71,0) \rightarrow 0.994$  & $(2.17,\frac{3\pi}{2}) \rightarrow 0.972$
& $(12.7,0) \rightarrow 0.984$ & $(62.2,0) \rightarrow 0.977$
& $(4.6,\frac{3\pi}{2}) \rightarrow 0.971$ \\
 & $\langle GS | Pf \rangle$ & 0.952 & 0.945 & 0.981 & 0.976 &
0.956 \\
& $\langle GS | APf \rangle$ & 0.611 & 0.842 & 0.164 & 0.315 &
0.598 \\
\hline
\end{tabular}
\caption{Comparing with the results in Table I in the main tex, the ground
states here are calculated by using perturbation theory only. The
overlaps between ground states and different trial wave functions and
the PH parity of the ground states.}\label{table2}
\end{table*}

We would like to compare different algorithms which are represented by the choices of the screening set $\mathcal{B}$. In Table \ref{table3}, we show the results for $N_e=6,\ldots 14$ systems obtained by using the screening theory only. This treatment obviously misses the PH symmetry breaking information by LLM even when $\kappa$ is not extremely large. The resulting ground states are PH symmetric. In Table \ref{table2}, we show the overlaps and the PH parities for $N_e = 6,\ldots 14$ systems calculated by the perturbation theory only, namely $\mathcal{B}=\emptyset$. In this case, the perturbation theory ignores the higher orders of $\kappa$ even though $\kappa$ is explicitly not a small quantity. The first order corrections of the Coulomb interaction tend the system Pfaffian-like \cite{pakrouski} and the wave function overlap $\langle GS | trial\rangle$ appears closer to unity ($>0.947$). However, when higher orders are taken into consideration, the properties of the system may become completely different.

\begin{table*}[htb]
\centering
\renewcommand\arraystretch{1.4}
\begin{tabular}{c|c|cccc}
\hline
 \multicolumn{2}{c|}{} & $N_e=6$ & $N_e=8$ & $N_e=10$ & $N_e=12$  \\
\hline
\multirow{4}*{\rotatebox{90}{\scriptsize{$(0,N_e/2 )$}}} &PH parity & 0.879 & 0.774 & 0.424 & 0.357 \\
&$(r,\theta) \rightarrow \langle GS | trial \rangle$
& $(7.66,0) \rightarrow 0.997$  & $(4.6,1.5\pi) \rightarrow 0.987$
& $(20.8,\pi) \rightarrow 0.990$ & $(17.2,\pi) \rightarrow 0.989$  \\
& $\langle GS | Pf \rangle$ & 0.994 & 0.974 & 0.989 & 0.987  \\
&$\langle GS | APf \rangle$ & 0.846 & 0.646 & 0.380 & 0.308  \\
\hline
\multirow{4}*{\rotatebox{90}{\scriptsize{$(N_e/2,N_e/2 )$}}} &PH parity  & 0.566 & 0.889 & 0.224 & 0.504 \\
&$(r,\theta) \rightarrow \langle GS | trial \rangle$
& $(7.12,0) \rightarrow 0.996$  & $(4.06,1.5\pi) \rightarrow 0.988$
& $(15.4,0) \rightarrow 0.993$ & $(10.0,\pi) \rightarrow 0.987$  \\
& $\langle GS | Pf \rangle$ & 0.989 & 0.978 & 0.990 & 0.983  \\
&$\langle GS | APf \rangle$ & 0.471 & 0.807 & 0.156 & 0.414  \\
\hline
\end{tabular}
\caption{
The overlaps between ground states and different trial wave
functions and the PH parity of the ground states, comparing with
the results in Table I in the main tex, the screening set is chosen as
$\mathcal{B}=\{ (1,0,\pm), (1,1,-),(2,0,\pm)\}$ here. }\label{table4}
\end{table*}

\begin{table*}[htbp]
\centering
\renewcommand\arraystretch{1.4}
\begin{tabular}{c|c|cccc}
\hline
\multicolumn{2}{c|}{} &  $N_e=6$ & $N_e=8$ & $N_e=10$ & $N_e=12$  \\
\hline
\multirow{4}*{\rotatebox{90}{\scriptsize{$(0,N_e/2 )$}}} &PH parity & 0.889 & 0.956 & 0.721 & 0.0113 \\
&$(r,\theta) \rightarrow \langle GS | trial \rangle$
& $(0.08,0) \rightarrow 0.804$  & $(0.64,1.5\pi) \rightarrow 0.801$
& $(0.6,0) \rightarrow 0.739$ & $(0.22,0) \rightarrow 0.698$  \\
& $\langle GS | Pf \rangle$ & 0.671 & 0.643 & 0.191 & 0.0351  \\
&$\langle GS | APf \rangle$ & 0.804 & 0.741 & 0.606 & 0.682  \\
\hline
\multirow{4}*{\rotatebox{90}{\scriptsize{$(N_e/2,N_e/2 )$}}} &PH parity & 0.576 & 0.962 & 0.625 & 0.367 \\
&$(r,\theta) \rightarrow \langle GS | trial \rangle$
& $(0.14,0) \rightarrow 0.993$  & $(0.44,1.5\pi) \rightarrow 0.808$
& $(0.34,\pi) \rightarrow 0.759$ & $(0.18,0) \rightarrow 0.555$  \\
& $\langle GS | Pf \rangle$ & 0.477 & 0.695 & 0.181 & 0.0938  \\
&$\langle GS | APf \rangle$ & 0.985 & 0.788 & 0.718 & 0.547  \\
\hline
\end{tabular}
\caption{The overlaps between ground states and different trial wave
functions and the PH parity of the ground states. The results here are
obtained by choosing $\mathcal{B}=\{ (1,1,-),(2,0,\pm), (1,2,\pm)\}$ in the
algorithm. }
\label{table5}
\end{table*}

The Pfaffian or anti-Pfaffian tendency depends on different algorithms. When different LLs are included in the screening set $\mathcal{B}$, the results could be quite different. Without any physical reason, just as the references, we show two examples in Table \ref{table4} and Table \ref{table5}, respectively. In Table \ref{table4} we set $\mathcal{B}=\{ (1,0,\pm), (1,1,-),(2,0,\pm)\}$ while $\mathcal{B}=\{ (1,1,-),(2,0,\pm), (1,2,\pm)\}$ in Table \ref{table5}. It is concluded that the LLs $(1,0,\pm)$ in perturbation theory mainly lead the system anti-Pfaffian state favored while the LLs $(1,2,\pm)$ mainly make the system tend to the Pfaffian state. We also note that the overlaps shown in Table \ref{table4} are very high $\sim 0.99$ in all the cases (the highest among all the algorithms), i.e. the ground state is almost described by the Pfaffian state. However, it does not necessarily mean that the ground state of $5/2$ FQHE is explained by the calculations since the set $\mathcal{B}$ is not chosen reasonably and the perturbation is only up to the first order.

\onecolumngrid
\section{The calculated ground states in a range of experimental parameters}

\begin{figure}[htbp]
\centering
\includegraphics[width=15cm]{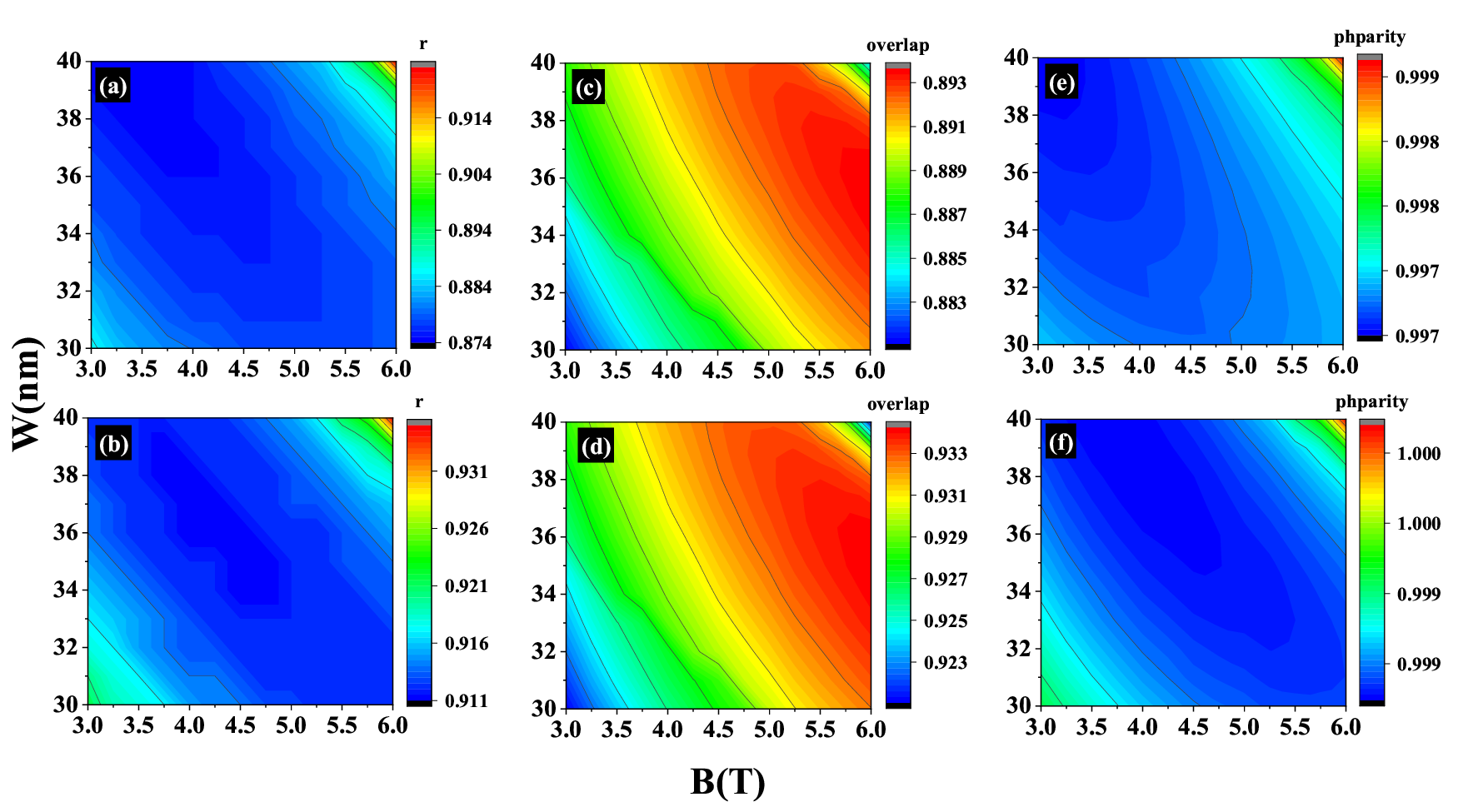}
\caption{(Color online) The ground state results plotted for the $N_e=8$ system with the experimental parameters varying in the range of $B\in[3,6]$ T and $W\in[30,40]$ nm. The optimal $r$ of the trial wave function at sectors (a) $(0,N_e/2)$ and (b) $(N_e/2,N_e/2)$. The overlaps between the ground state and the optimized trial wave function at sectors (c) $(0,N_e/2)$ and (d) $(N_e/2,N_e/2)$. The particle-hole parities of the ground states at sectors (e) $(0,N_e/2)$ and (f) $(N_e/2,N_e/2)$.}
\label{figS2}
\end{figure}

\begin{figure}[htbp]
\centering
\includegraphics[width=15cm]{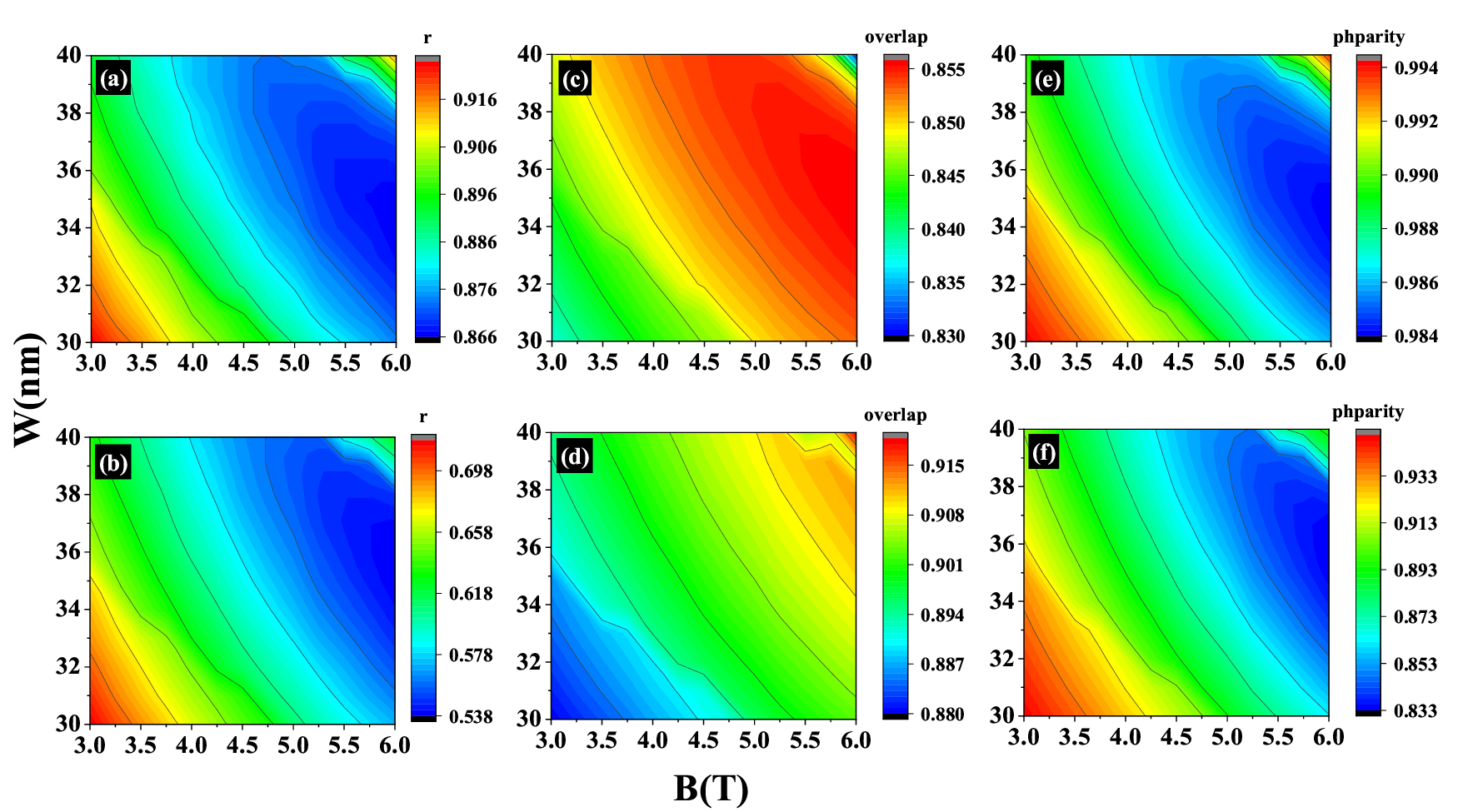}
\caption{(Color online) The same plots as what are shown in Fig. \ref{figS2}. The electrons number of the system here is $N_e=10$.
}
\label{figS3}
\end{figure}

\begin{figure}[htbp]
\centering
\includegraphics[width=15cm]{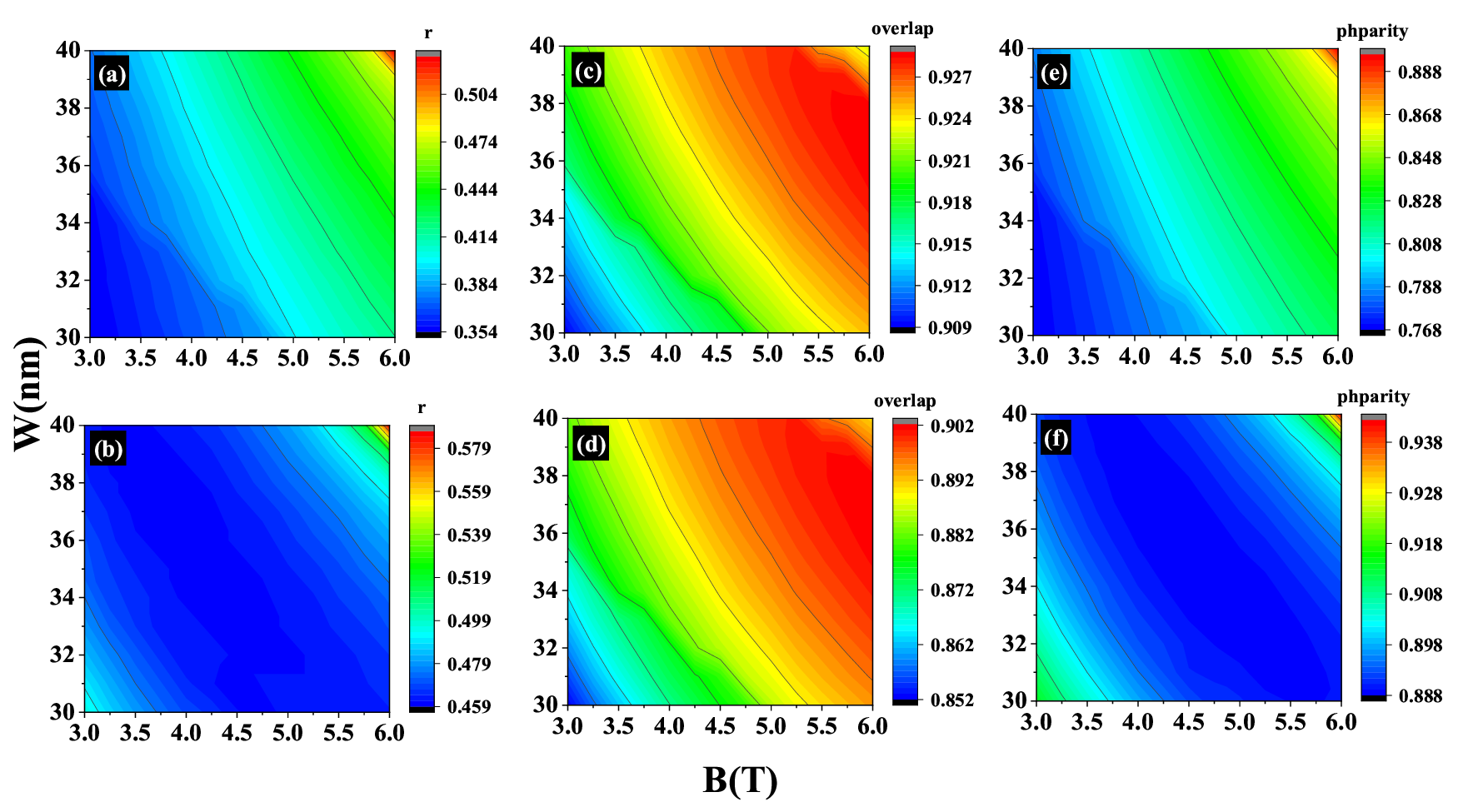}
\caption{(Color online) The same plots as plotted in Fig. \ref{figS2}. The electrons number of the system here is $N_e=12$.
}
\label{figS4}
\end{figure}

\begin{figure}[htbp]
\centering
\includegraphics[width=15cm]{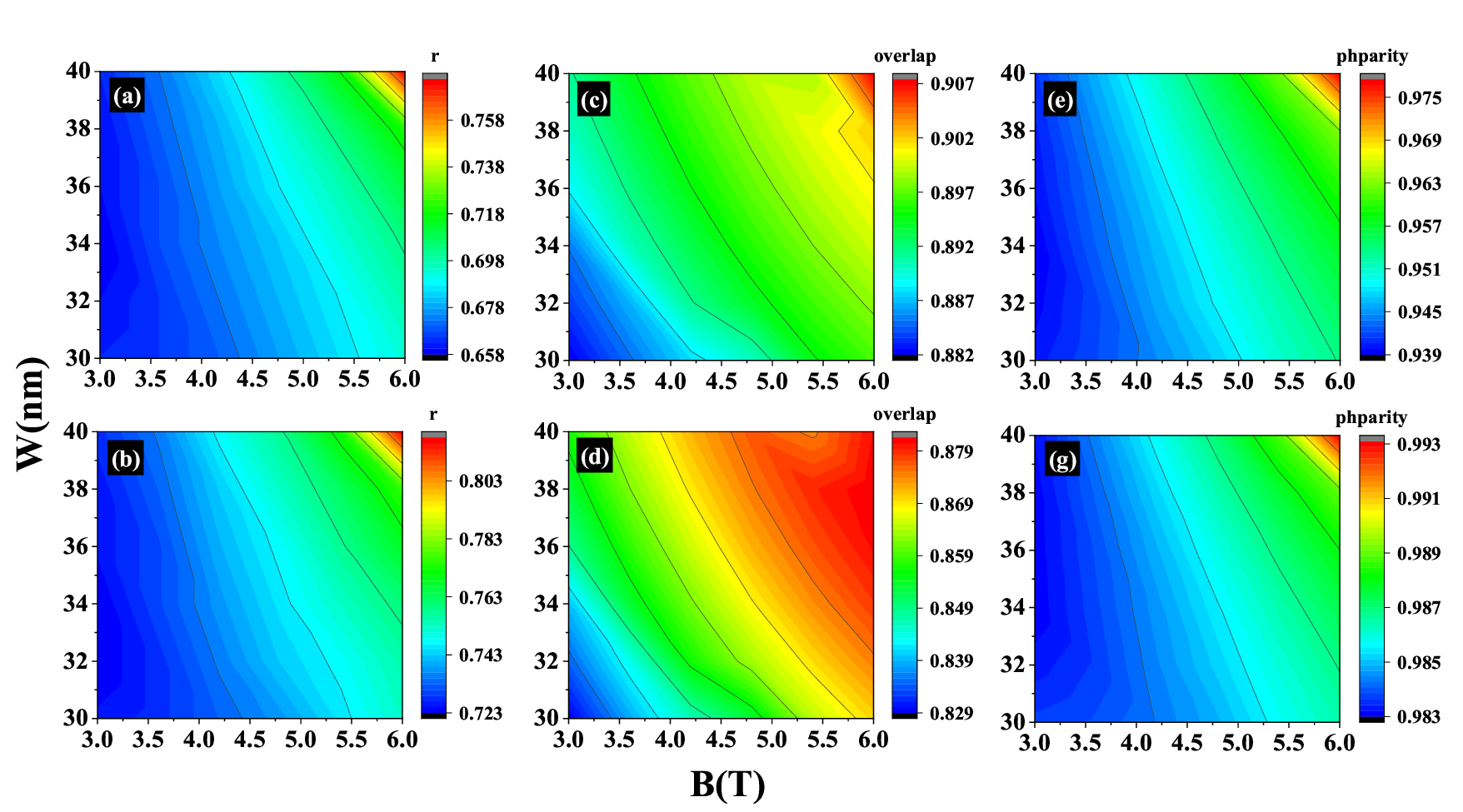}
\caption{(Color online) The same plots as what are plotted in Fig. \ref{figS2}. The electrons number of the system is $N_e=14$.
}
\label{figS5}
\end{figure}

Considering the experimental environment, we study the ground states of the systems in the parameter range of $B\in[3,6]$ T and $W\in[30,40]$ nm. The optimal ratio $r$ in the trial wave function, the ground state overlap with the optimized trial wave function, and the PH parity of the ground state as functions of $B$ and $W$ parameters are shown in the contour plots of Figs. \ref{figS2} to \ref{figS5} for systems with $N_e=8$ to $14$, respectively. In general, the PH symmetry of the ground states are weakly broken with their PH parity in the vicinity of unity. The ground states can be well described by the variational trial wave function with a larger overlap close to unity in the full parameter range. The value of the optimal $r$ varies with the external parameters but remains less than $1$, indicating the anti-Pfaffian component more favorable in the ground state.

\onecolumngrid
\section{Entanglement spectra for even-number electron systems}

\begin{figure}[htbp]
\centering
\includegraphics[width=9.5cm]{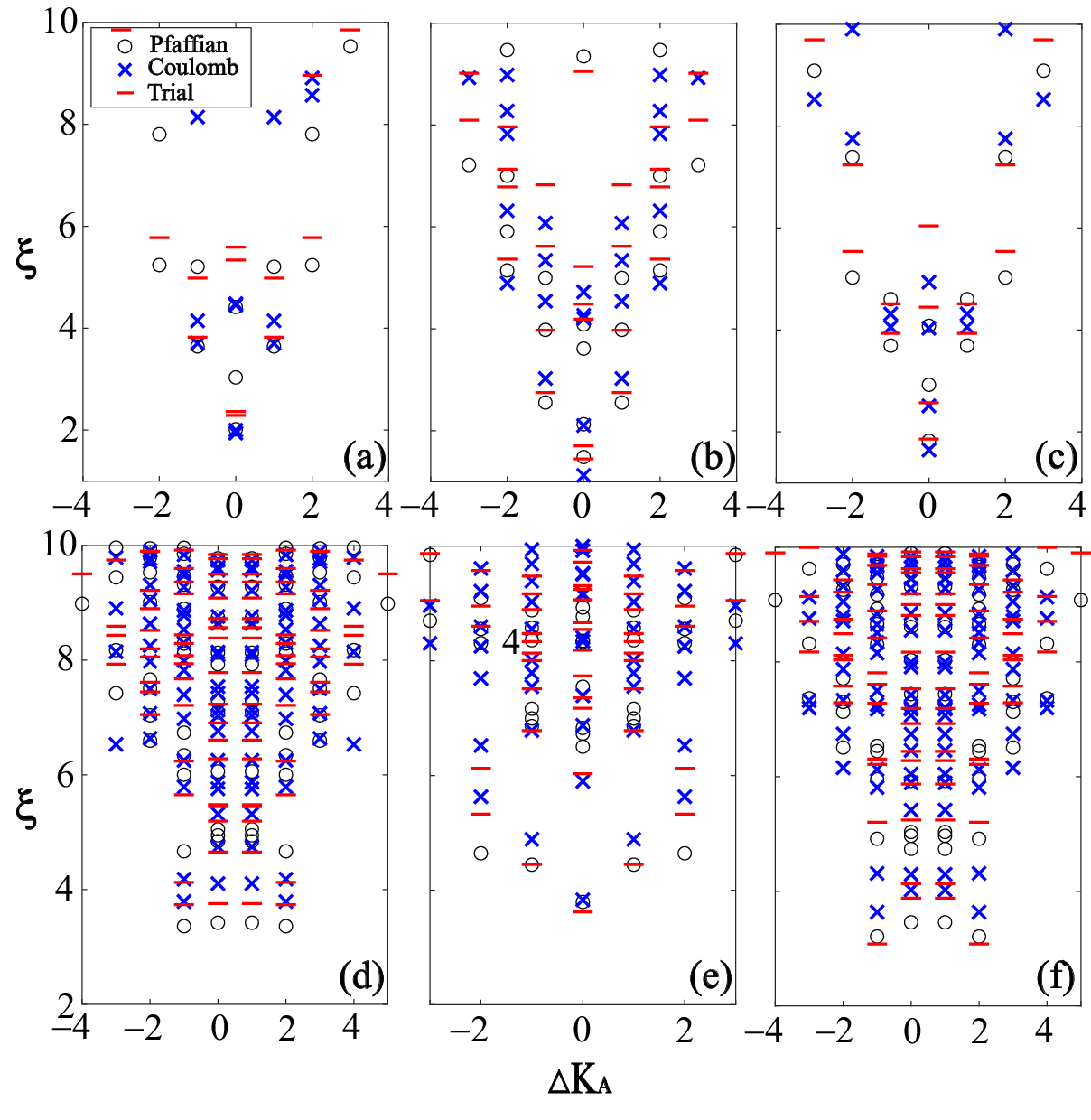}
\caption{(Color online) The entanglement spectra of the Coulomb ground states, Pfaffian states and optimal trial wave functions for $N_e = 8$ at
pseudomomentum sectors (a) $(0,4)$, (b) $(4,0)$ and (c) $(4,4)$, and for $N_e = 14$ at sectors (d) $(0,7)$, (e) $(7,0)$ and (f) $(7,7)$, respectively. The bipartition blocks are evenly cut with the half number of electrons and orbits at each block. For comparison and simplicity, only the major tower around the lowest level is shown in each spectrum.
}
\label{fig3}
\end{figure}

\begin{figure}[htbp]
\centering
\includegraphics[width=9.5cm]{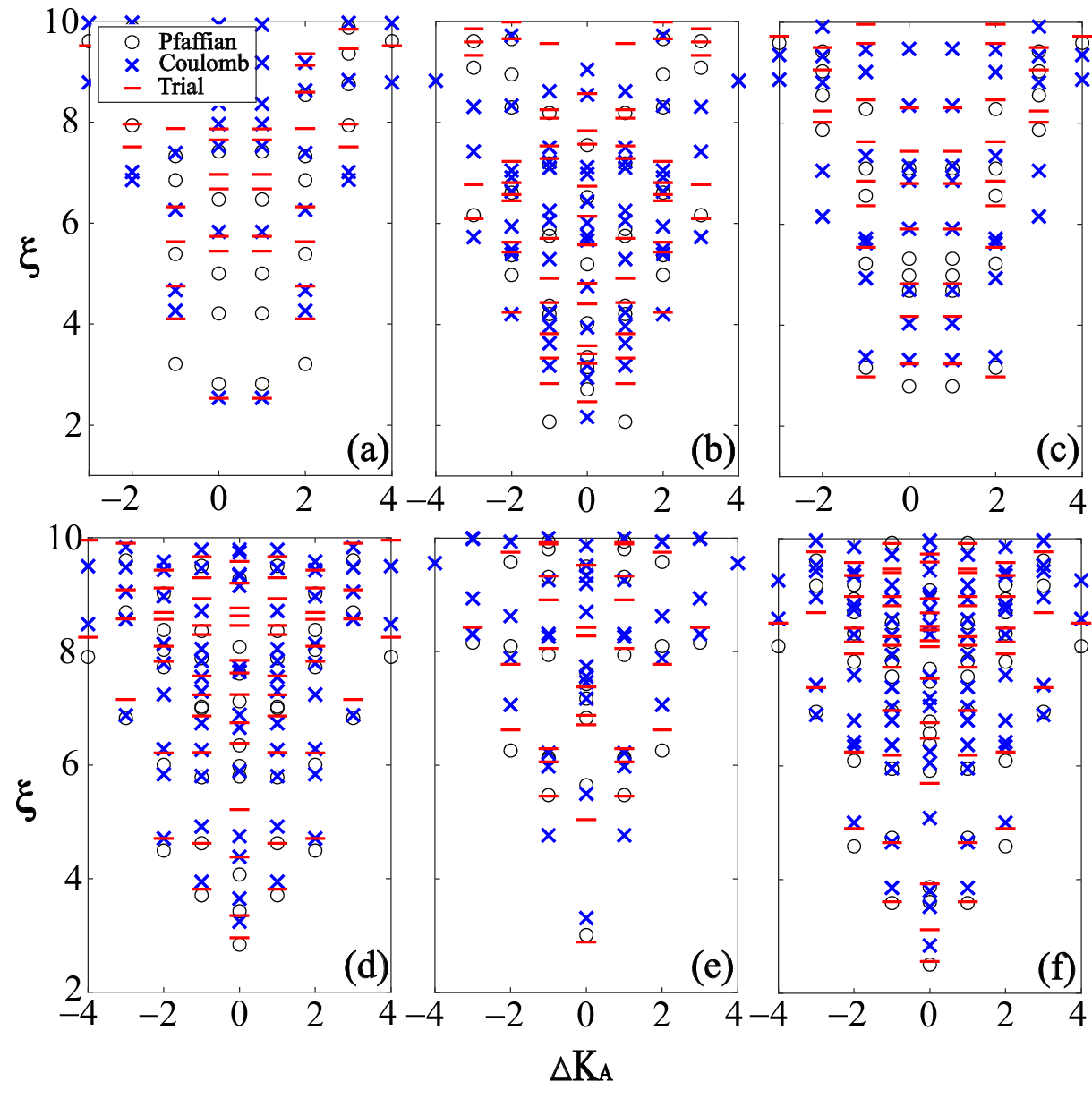}
\caption{(Color online) Similar with Fig. \ref{fig3}, the entanglement spectra are shown for $N_e = 10$ at pseudomomentum sectors (a) $(0,5)$, (b) $(5,0)$ and (c) $(5,5)$, and for $N_e = 12$ at  sectors (d) $(0,6)$, (e) $(6,0)$ and (f) $(6,6)$. The bipartition blocks are evenly cut and only the major tower is shown.
}
\label{fig3b}
\end{figure}

To explore the topological nature of the ground state, we calculate its ES on torus with even bipartition cutting on LL orbits for several even-electron example systems. The cutting on torus creates two chiral edges. Around the cell geometry $|\tau|\sim 1$, these two edges interact with each other and it is difficult to resolve either edge alone. Nevertheless, we can compare the ES structures of the low-lying levels in the ground state with those in the candidate wave functions to gain more understanding. In Figs. \ref{fig3} and \ref{fig3b}, we exhibit ES of the ground state, the Pfaffian (anti-Pfaffian) state, and the optimal trial wave function for $N_e=8,10,12,14$ systems. On torus, an anti-Pfaffian state has the same ES as its corresponding Pfaffian state. Compared with the Pfaffian (or anti-Pfaffian) state, the optimized trial wave function generally shows a better match to the ground state in the low-lying ES, reconfirming it as a better candidate. A clearer ES containing more information on the edge excitations of the conformal field theory may await the future studies with a cylindrical geometry \cite{zhu}.

\end{document}